# Unraveling the Effect of Circularly Polarized Light on Reciprocal Media: Breaking Time Reversal Symmetry with Non-Maxwellian Magnetic-esque Fields

R. Merlin

*The Harrison M. Randall Laboratory of Physics,*
*University of Michigan, Ann Arbor, Michigan 48109-1040, USA*

Optical rectification of intense, circularly polarized light penetrating a material generates a static magnetic field aligned with the light's direction and proportional to its intensity. Recent experiments have unveiled a substantial, orders-of-magnitude gap between the observed magnetization and theoretical predictions. In this study, we demonstrate that circularly polarized light creates large non-Maxwellian fields that disrupt time-reversal symmetry, effectively mimicking authentic magnetic fields within the material while eluding detection externally. These unconventional fields give rise to Faraday-rotation-like phenomena, which are a high-frequency manifestation of Berry's curvature.

The discovery of light-induced magnetization, known as the inverse Faraday effect (IFE), has a long history dating back many years [1]. When high-intensity, circularly polarized radiation propagates through a material, it generates a static magnetic field through optical rectification. This field aligns itself with the direction of light propagation, its magnitude is directly proportional to the intensity of the incident beam and, as in the conventional Faraday effect, it can be examined by measuring the rotation in the linear polarization of a second, copropagating beam. Recent experiments have unveiled a striking discrepancy: the measured values of the light-induced magnetization are orders of magnitude greater than what theoretical predictions [2,3,4], based on our current understanding of the IFE, would suggest [5,6,7,8,9]. In this study, we demonstrate that apart from the IFE-magnetization, circularly polarized radiation generates unconventional non-Maxwellian fields that lift Kramers' degeneracy, and, for all practical purposes, they perfectly mimic the behavior of a genuine magnetic field within the material. However, unlike true magnetic fields, these fields remain undetectable outside. Since these fields do not depend on magnetic coupling, their strength significantly surpasses the predictions of the IFE [10].

Given its relevance to the discussion, we provide a brief overview of the direct and inverse Faraday effects. For simplicity, we consider a material that is non-magnetic and belongs to the cubic system. Following the Armstrong-Bloembergen-Ducuing-Pershan (ABDP) approach [11], the interaction between a medium and the electromagnetic field is described by a phenomenological free energy density, which depends only on the set of Fourier components of the electric field $\{e_k(\omega_n)\}$, where $e_k(t) = \sum_n e_k(\omega_n)\exp(i\omega_n t)$ with $e_k(\omega_n) = e_k^*(-\omega_n)$ and $k = x, y, z$. The Faraday free energy can be written as [12]

$$F = -\sum_{jk,l} \chi^F_{jk,l} e_j^*(\omega) e_k(\omega) H_l \qquad (1)$$

where **H** is a static magnetic field and $\chi^F$ is the Faraday susceptibility tensor (not to be confused with the conventional nonlinear susceptibility $\chi^{(2)}$). Since $F$ must be invariant under time reversal, this tensor is anti-symmetric with respect to indices $(j,k)$ and, moreover, its components are all imaginary for lossless media (note that time reversal maps $i$ to $-i$). Thus, for $\mathbf{H}=(0,0,H_0)$ and a monochromatic wave of frequency ω with its wavevector parallel to the field,

$$F = -H_0 \chi^F_{xy,z}\left(e_x e_y^* - e_y e_x^*\right) = iH_0 \chi^F_{xy,z}\left(e_{R,z} e_{R,z}^* - e_{L,z} e_{L,z}^*\right) \tag{2}$$

where $e_{R,z} = (e_x - ie_y)2^{-1/2}$ and $e_{L,z} = (e_x + ie_y)2^{-1/2}$ are the circular polarization eigenvectors. Since the polarization is given by $p_k = -\partial F/\partial e_k^*$, this expression adds an antisymmetric component to the permittivity

$$\delta\varepsilon_{xy} = -\delta\varepsilon_{yx} = 4\pi i \left|\chi^F_{xy,z}\right| H_0 \tag{3}$$

which accounts for the circular birefringence and concomitant polarization rotation (Faraday effect). Moreover, absent the field and under excitation with circularly polarized light, Eq. (2) predicts the generation of a static IFE magnetization along the $z$ direction

$$M_z = -\frac{\partial F}{\partial H_0} = \left|\chi^F_{xy,z}\right|\left[\mathcal{E}_{R,z}(\Omega)\mathcal{E}_{R,z}^*(\Omega) - \mathcal{E}_{L,z}(\Omega)\mathcal{E}_{L,z}^*(\Omega)\right]. \tag{4}$$

Here, the notation has been altered from lowercase to uppercase letters to underscore the distinction: the field serves as a probe for the actual magnetic field in the Faraday effect, whereas it functions as a pump, with components $E_k(t) = \sum_n \mathcal{E}_k(\Omega_n)\exp(i\Omega_n t)$, in the IFE (throughout, we use Ω and ω to indicate, respectively, the pump and probe frequencies). Experiments have shown that this magnetization is a proper Maxwellian field that can be probed by a pickup coil placed outside the sample [1].

The static bilinear combination $\mathcal{E}_{R,z}\mathcal{E}_{R,z}^* - \mathcal{E}_{L,z}\mathcal{E}_{L,z}^*$ is the $z$-component of the vector product $i(\vec{\mathcal{E}} \times \vec{\mathcal{E}}^*)$. As such, it transforms like a magnetic field since it is even under inversion and odd under time reversal. Thus, it is apparent that coupling of such a combination to the probe field will lead to a rotation of the probe polarization that is indistinguishable from the Faraday effect. More precisely, for fields that lie in the $xy$ plane, the induced probe polarization resulting from a free energy of the form

$$F = \Lambda\left[\vec{\mathcal{E}}(\Omega) \times \vec{\mathcal{E}}^*(\Omega)\right] \cdot \left[\vec{e}(\omega) \times \vec{e}^*(\omega)\right] \tag{5}$$

is

$$\begin{aligned}
p_x(\omega) &= -\partial F / \partial e_x^* = +\Lambda\left[\mathcal{E}_x(\Omega)\mathcal{E}_y^*(\Omega) - \mathcal{E}_y(\Omega)\mathcal{E}_x^*(\Omega)\right]e_y(\omega) \\
p_y(\omega) &= -\partial F / \partial e_y^* = -\Lambda\left[\mathcal{E}_x(\Omega)\mathcal{E}_y^*(\Omega) - \mathcal{E}_y(\Omega)\mathcal{E}_x^*(\Omega)\right]e_x(\omega)
\end{aligned} \tag{6}$$

where $\Lambda$ is a real constant whose value is estimated below. Provided $\mathcal{E}_x(\Omega)\mathcal{E}_y^*(\Omega) \neq \mathcal{E}_y(\Omega)\mathcal{E}_x^*(\Omega)$ and $\Lambda \neq 0$, the antisymmetric component in Eq. (6) embodies a non-Maxwellian field, which makes the material non-reciprocal by emulating the effect of a static magnetic field through optical rectification. It is important to underscore that the rotation cannot be explained by chiral optical activity, which is reciprocal. Moreover, the induced magnetization is here zero because the free energy of Eq. (5) does not depend on the magnetic field. Therefore, in contrast to the IFE, the pseudo magnetic field is not detectable beyond the sample's boundaries. When we consolidate all these factors, it becomes evident that the pump bilinear combination serves two distinct functions. On one hand, it generates magnetization as described in Eq. (4). On the other hand, it induces an unrelated rotation in the polarization of a linearly polarized probe, effectively emulating the Faraday effect. Albeit not extensively, the latter phenomenon has been previously examined in the theoretical literature, where it is referred to as the optical Faraday effect [13,14].

In order to provide an estimate of the parameter Λ, consider the total Hamiltonian $\hat{H} = \hat{H}_0 + \mathbf{E}.\vec{\Pi}$ where $\hat{H}_0$ describes the isolated system, with eigenenergies $\hbar\omega_p$, and

$$\mathbf{E}.\vec{\Pi} = e(r_+ \mathcal{E}_R + r_- \mathcal{E}_L)e^{-i\Omega t} + c.c. \tag{7}$$

is the electric-dipole coupling, with $r_\pm = (x \pm iy)2^{-1/2}$. Our focus lies in examining the consequences stemming from the rectification of the perturbation. Specifically, we are concerned with terms quadratic on $\mathbf{E}.\vec{\Pi}$ that combine positive and negative frequencies, which result in the emergence of the static field. Using standard time-dependent perturbation methods, we obtain the antisymmetric mixing term [15]

$$\langle q|\hat{H}_{AS}|p\rangle = \frac{e^2}{\hbar}\left(\mathcal{E} \times \mathcal{E}^*\right) \cdot \sum_n \frac{\Omega}{(\omega_n - \omega_p)^2 - \Omega^2}\left(\mathbf{r}_{qn} \times \mathbf{r}_{np}\right) , \tag{8}$$

which describes the effect of the pump on the states of the system, ultimately leading to the breakdown of time-reversal invariance. We observe that although spin-orbit coupling is not essential for generating a magnetization, its presence plays a crucial role in lifting Kramers' degeneracy for, in its absence, the quantum states are real and, thus, $\mathbf{r}_{qn} \times \mathbf{r}_{np} = 0$. An example of spin-orbit induced mixing of Kramers' doublets is shown in Fig. 1.

The pump-induced magnetization breaks Kramers' degeneracy directly by coupling to the electron magnetic moment. The effective mixing term is

$$\langle q|\hat{H}_M|p\rangle = \mu_B \left|\chi^F_{xy,z}\right|\left(\mathcal{E} \times \mathcal{E}^*\right) \cdot (\mathbf{l}_{qp} + 2\mathbf{s}_{qp}) , \tag{9}$$

where $\mathbf{s}$ and $\mathbf{l}$ are the electron spin and angular momentum, and $\mu_B$ is the Bohr magneton. To facilitate a comparison with Eq. (8), we employ the low-frequency limit of the electron contribution to the Faraday susceptibility, $\chi^F_e(\Omega) \approx \Omega\mu_B / \hbar\pi\Delta_G^2$, specifically applicable to transparent media [16]; $\Delta_G$ denotes the bandgap frequency. By doing so, we derive an order-of-magnitude estimate that establishes a relationship between the ratio of mixing terms and the ratio of fundamental electric and magnetic dipole moments

$$\left|\frac{\langle q|\hat{H}_{AS}|p\rangle}{\langle q|\hat{H}_{M}|p\rangle}\right| \sim \left|\frac{\Lambda}{\chi^{F}}\right| > \left(\frac{ea_0}{\mu_B}\right)^2 = 4\alpha^{-2} \approx 7.5 \times 10^4 \quad . \tag{10}$$

Here, $a_0$ is the Bohr radius and $\alpha$ is the fine structure constant. This key result explains the substantial disparity in magnitude between the pseudo-magnetic field and the IFE magnetization. Since the effective mass for vibrations significantly surpasses the electron mass, the corresponding ratio is considerably larger for phonons [19,20,21,22].

The above discussion is evidently applicable not only to **E** but extends to any coherent field **Q** that carries an electric dipole, such as the field of an IR-active phonon or a dipole-allowed exciton, as well as to the cross product of two arbitrary vector fields. As for the electric field, coupling the probe field to the bilinear combination $\mathcal{Q}_x(\Omega)\mathcal{Q}_y^*(\Omega) - \mathcal{Q}_y(\Omega)\mathcal{Q}_x^*(\Omega)$, where $\{\mathcal{Q}_k(\Omega)\}$ is the Fourier set derived from **Q**(*t*), leads to a Faraday-like rotation of the probe polarization. The **Q**-field breaks time-reversal invariance by coupling to electron states through an interaction of the form $\vec{\Xi}.\mathbf{Q}$ where the vector operator $\vec{\Xi}$ is the deformation potential interaction for vibrational modes and the electron-electron interaction for excitons. The corresponding mixing term can be obtained from Eq. (8) by replacing $e^2(\mathcal{E}\times\mathcal{E}^*).(\mathbf{r}_{qn}\times\mathbf{r}_{np})$ with $(\mathcal{Q}\times\mathcal{Q}^*).(\vec{\Xi}_{qn}\times\vec{\Xi}_{np})$.

In accordance with the established framework of nonlinear optics (NLO) [17], the coupling of the probe field to doubly bilinear combinations involving, say, the vector fields **U** and **V,** can be interpreted as describing one of the numerous four-wave-mixing processes linked to the generalized free energy $\sum_{klmn}\beta_{klmn}\mathcal{U}_k\mathcal{V}_l^*e_m e_n^*$, where $\beta^{(3)}$ is a third-order nonlinear susceptibility, which possesses the same symmetry properties as the optical susceptibility $\chi^{(3)}$, but has a very different origin. As opposed to $\chi^{(3)}$, for which all the terms are quartic in the field, this free energy may contain terms that are both quadratic in the electric field and in **Q**, or cubic in the field and linear in **Q**, as for the stimulated hyper-Raman effect. Regardless, instead of delving into the symmetry properties of the tensor, as in standard NLO, it proves more advantageous for our discussion to consider an expansion that separates the different irreducible representations of the point group. Directing our attention towards dipole-carrying excitations, the most general form of the free energy density for fields in the *xy* plane is

$$F = C(A)(Q_x Q_x^* + Q_y Q_y^*)(e_x e_x^* + e_y e_y^*) +$$
$$C(E)\left[\left(Q_x Q_y^* + Q_y Q_x^*\right)\left(e_x e_y^* + e_y e_x^*\right) - 3^{1/2}(Q_x Q_x^* - Q_y Q_y^*)(e_x e_x^* - e_y e_y^*)\right] + \qquad (11)$$
$$C(T_S)\left(Q_x Q_y^* + Q_y Q_x^*\right)\left(e_x e_y^* + e_y e_x^*\right) + C(T_{AS})(Q_x Q_y^* - Q_y Q_x^*)(e_x e_y^* - e_y e_x^*)$$

where $C(A)$, $C(E)$, $C(T_S)$ and $C(T_{AS})$ are frequency-dependent coefficients associated, respectively, with the fully symmetric, the doubly degenerate and the triply degenerate symmetric ($T_S$) and anti-symmetric ($T_{AS}$) representations of the cubic system. Certainly, the same expansion applies to **E**. Note that the well-known symmetry arguments that rule out the antisymmetric contribution for the electric field and, thus, for **Q**, particularly Kleinman's rules [18] do not apply if time-inversion symmetry is broken. In the case of $\chi^{(3)}$, the symmetric terms are associated with the optical Kerr effect, which describes nonlinear changes in the refractive index that are proportional to the pump intensity [17]. In this context, the anti-symmetric term for **E** represents the magnetic counterpart to the symmetric Kerr effect. Since **Q** couples linearly to **E**, that is, $Q_k(\Omega) = \chi_Q \mathcal{E}_k(\Omega)$, where $\chi_Q$ is proportional to $\chi^{(1)}$, both, the optical Kerr and the pseudo-IFE will be enhanced when the pump frequency is tuned to resonate with the natural frequency of these modes.

The experiments we wish to examine are all of the pump-probe type. The measurements of Basini et al. [5] on SrTiO3 use resonant, circularly-polarized THz pulses to bring the soft mode into circular motion and a 400 nm linearly polarized probe to measure the Faraday rotation resulting from the pump-induced magnetization. Their results show that the magnetization inferred from Eq. (4) is a factor of $\approx 10^4$ smaller than the experimental value. Similarly, the work of Mikhaylovskiy et al. [6] on Tb3Ga5O12 using, both, 800-nm pump and probe pulses, finds a four orders of magnitude discrepancy between the theoretical Verdet constant and the measurements. The observation that circularly polarized pulses tuned to resonate with IR-active phonons leads to large coherent spin precessions in ErFeO3 [7] and in DyFeO3 [8] further amplifies skepticism to the idea that the IFE mechanism alone fully explains the phenomenon of light-induced magnetization. This also applies to the studies of Kim et al. [9] who reported effective magnetic fields as large as 60 T in experiments on WeS2 monolayers using circular excitation below but near the *A*-exciton resonance.

In order to provide a quantitative comparison of the diverse mechanisms responsible for polarization rotation, one needs reliable estimates for $\chi^F$ and the anti-symmetric factor $C(T_{AS})$. For the electronic contribution to the Faraday susceptibility, we use Becquerel's formula [16]

$$\chi_e^F(\omega) = \frac{Ce}{4\pi m_0 c} \frac{\omega \Delta_G^2}{(\Delta_G^2 - \omega^2 - i\Gamma\omega)^2} \quad , \tag{12}$$

which applies to transparent, diamagnetic media at frequencies ω below the bandgap. Here, $\Delta_G$ and $\Gamma$ are the bandgap frequency and the associated decay rate, *C* is a constant of order one, $m_0$ and *e* are the mass and charge of the electron, and *c* is the speed of light. The expression for the phonon counterpart, $\chi_p^F$, applies to substances with a doubly- or triply degenerate IR-active vibration (non-degenerate modes do not contribute to the magnetic moment). It is of the same form and can be obtained from Eq. (12) by replacing $\Delta_G$, $m_0$ and $e$ with, respectively, the phonon frequency, $\Omega_0$, the effective mass and the Born charge of the mode, together with the substitution $C \rightarrow (\varepsilon_0 - \varepsilon_\infty)$, where $\varepsilon_0$ ($\varepsilon_\infty$) is the static (high-frequency) dielectric constant [19,20].

The coefficients in Eq. (11) and, in particular, $C(T_{AS})$ can be written as linear combinations of the cartesian elements of $\beta^{(3)}$, which can be obtained from standard calculations of the induced polarization using time-dependent perturbation theory [17]. Instead of finding the right combination to perform the calculations, however, it is significantly simpler to obtain $C(T_{AS})$ directly from Eq. (8), given that the primary effect of the symmetry-breaking term is the splitting of Kramers' doublets, which is in the end the underlying cause of circular birefringence. A crude estimate gives

$$C(T_{AS}) \sim \omega \frac{d\chi^{(1)}}{d\omega} \frac{\tilde{\Xi}^2 a_L^6}{\hbar^2 \Delta_G^2} \quad . \tag{13}$$

Here, $\chi^{(1)}$ is the first-order optical susceptibility and $a_L$ is the lattice parameter. As before, $\vec{\Xi}$ is the deformation potential coupling and $\omega$ is the probe frequency. The tilde mark denotes an average over all the states. For insulators, using that $\chi^{(1)} \approx \tilde{\Pi}^2 / (\Delta_G - \omega)$, $\hbar\Delta_G = 2$ eV, $a_L = 5$ Å and the very conservative values of $\tilde{\Xi} = 1$ V/Å and $\tilde{\Pi} = 0.1$ V/Å for the couplings, we obtain from Eq. (13) $C(T_{AS}) \approx (\omega/\Delta_G) 6 \times 10^{-21} \text{m}^2/\text{V}^2$ for $\omega \ll \Delta_G$. The factor is about 2-3 orders of magnitude larger than typical off-resonance values for $\chi^{(3)}$, reflecting the fact that the ratio between the deformation potential and electric dipole interaction is typically large.

To compare with the IFE, we write Eq. (11) in terms of the electric field by means of the replacement $\mathcal{Q}_k(\Omega) = \chi_Q \mathcal{E}_k(\Omega)$. Thus, for $\chi^{(1)} \approx \tilde{\Pi}^2 / (\Delta_G - \omega)$, the ratio between the non-Maxwellian and the IFE magnetization is

$$\left|\frac{M_{NM}}{M_{IFE}}\right| \approx \frac{|\chi_Q(\Omega)|^2}{\chi^F(\Omega)\chi^F(\omega)} \frac{\omega\Delta_G}{(\Delta_G - \omega)^2} \frac{\tilde{\Xi}^2\tilde{\Pi}^2 a_L^9}{\hbar^3 \Delta_G^3} \quad . \tag{14}$$

Using the phonon equivalent to Eq. (12) for $\chi^F(\Omega)$, probe frequencies $\omega \gg \Omega_0$ and Becquerel's formula for $\chi^F(\omega)$, we obtain $|M_{NM}/M_{IFE}| \sim 10^6$, a value that is larger than but consistent with the results of Basini et al. [5]. Parenthetically, we note that their observations of a signal at twice the pump frequency can be accounted for by the field $\vec{\mathcal{C}}(\Omega) \times \vec{\mathcal{E}}(\Omega)$ since $\vec{\mathcal{C}} \perp \vec{\mathcal{E}}$ near resonance. In addition, for $\omega \approx \Delta_G/2$ and $\tilde{\Theta} \sim 1$ V/Å, one gets $|M_{NM}/M_{IFE}| \sim 5\times 10^3$, which is in fairly good agreement with the measurements at $\omega = \Omega \gg \Omega_0$ reported by Mikhaylovskiy et al. [6]. Concerning the near-resonant studies of Kim et al. [9], it is worth mentioning that Eq. (14) predicts an enhancement of $\sim 6\times 10^3$ for a detuning of 0.05 eV. Ultimately, the large enhancements vis-à-vis the conventional IFE stem from the fact that our approach relies on the interaction with the electric field (through $\Pi$) or on electrostrictive-like effects (through $\Xi$) whereas the IFE reflects the much weaker magnetic coupling. Finally, we note that mechanisms involving phonon-mediated changes in, say, the crystal field or exchange constants in magnetic crystals can generally be framed in terms of a coupling similar to that proposed here [21,22,23,24]. This makes it possible for a non-Maxwellian field to couple to spins much in the same way a real magnetic field does, leading to light-induced coherent spin precessions, as reported in Ref. [7] and Ref. [8].

In closing, we demonstrate a connection between non-Maxwellian fields and Berry's curvature. Instead of the dipolar interaction, we change gauge and write the coupling of the electrons to the probe field as $(e/m_0 c)\sum \mathbf{p}_l \cdot \mathbf{A}(t) = -i(e/m_0\omega)\sum \mathbf{p}_l \cdot \mathbf{E}(t)$ where $\mathbf{p}$ is the electron momentum operator. Thus, Eq. (8) is replaced by

$$\langle q|\hat{H}_{AS}|p\rangle = -\frac{e^2}{\hbar m_0^2 \Omega}(\mathcal{E}\times\mathcal{E}^*)\cdot\sum_n \frac{(\mathbf{p}_{qn}\times\mathbf{p}_{np})}{(\omega_n-\omega_p)^2-\Omega^2} \quad . \tag{15}$$

Provided the light wavelength is much larger than the lattice parameter, this term preserves periodicity and, thus, the eigenstates of the perturbed Hamiltonian maintain the Bloch form.

The difference between the off-diagonal components of the susceptibility is

$$\chi_{xy}(\omega)-\chi_{yx}(\omega) = \left(\frac{e}{m_0\omega}\right)^2 \sum_{n\in vb,\, m\in cb, \mathbf{k}} \left[ \frac{\langle\Phi_{n\mathbf{k}}|p_y|\Phi_{m\mathbf{k}}\rangle\langle\tilde{\Phi}_{m\mathbf{k}}|p_x|\Phi_{n\mathbf{k}}\rangle}{\hbar(\omega_{mn,\mathbf{k}}-\omega)} - \frac{\langle\Phi_{n\mathbf{k}}|p_x|\Phi_{m\mathbf{k}}\rangle\langle\Phi_{m\mathbf{k}}|p_y|\Phi_{n\mathbf{k}}\rangle}{\hbar(\omega_{mn,\mathbf{k}}-\omega)} \right] . \tag{16}$$

Here, $n$ and $m$ label states in the valence and conduction band, and $|\Phi_{s\mathbf{k}}\rangle$ is the periodic part of the exact eigenstate of the perturbed, symmetry-broken Hamiltonian, characterized by wavevector $\mathbf{k}$, band $s$ and energy $\hbar\omega_{s\mathbf{k}}$, and $\omega_{mn,\mathbf{k}} = \omega_{m\mathbf{k}}-\omega_{n\mathbf{k}}$. Using the $\mathbf{k}\cdot\mathbf{p}$ method, one can write $|\Phi_{s\mathbf{k}}\rangle$ as a superposition of states of wavevector $\mathbf{k}_0$

$$|\Phi_{s\mathbf{k}}\rangle \approx |\Phi_{s\mathbf{k}_0}\rangle + \frac{1}{m_0}\sum_{p\neq s}\frac{\langle\Phi_{s\mathbf{k}_0}|(\mathbf{k}-\mathbf{k}_0)\cdot\mathbf{p}|\Phi_{p\mathbf{k}_0}\rangle}{\hbar\omega_{sp,\mathbf{k}}}|\Phi_{p\mathbf{k}_0}\rangle \tag{17}$$

where $\mathbf{k}_0$ is a wavevector near $\mathbf{k}$. Therefore,

$$\nabla_{\mathbf{k}}|\Phi_{s\mathbf{k}}\rangle = \frac{1}{m_0}\sum_{p\neq s}\frac{\langle\Phi_{s\mathbf{k}}|\mathbf{p}|\Phi_{p\mathbf{k}}\rangle}{\hbar\omega_{sp}}|\Phi_{p\mathbf{k}}\rangle \tag{18}$$

and, thus, the $z$-component of the Berry curvature [25] can be written as

$$\Upsilon(s,\mathbf{k}) = \langle\nabla_{k_y}\Phi_{s\mathbf{k}}|\nabla_{k_x}\Phi_{s\mathbf{k}}\rangle - \langle\nabla_{k_x}\Phi_{s\mathbf{k}}|\nabla_{k_y}\Phi_{s\mathbf{k}}\rangle = \\ \frac{1}{m_0^2}\sum_{p\neq s}\frac{\langle\Phi_{s\mathbf{k}}|p_x|\Phi_{p\mathbf{k}}\rangle\langle\Phi_{p\mathbf{k}}|p_y|\Phi_{s\mathbf{k}}\rangle - \langle\Phi_{s\mathbf{k}}|p_y|\Phi_{p\mathbf{k}}\rangle\langle\Phi_{p\mathbf{k}}|p_x|\Phi_{s\mathbf{k}}\rangle}{\hbar^2\omega_{sp}^2} \quad . \tag{19}$$

Consider now a two-band problem and assume that intra-band dipole transitions are forbidden. Then, the comparison with Eq. (16) gives

$$\left.\frac{\partial(\chi_{xy}-\chi_{yx})}{\partial\omega}\right|_{\omega=0} \propto \sum_{s\mathbf{k}} \Upsilon(s,\mathbf{k}) \quad . \tag{20}$$

This expression sheds light on the relationship between light-induced time-reversal symmetry breaking and Berry's curvature, as well as on the emergence of the antisymmetric component of the susceptibility as its high frequency manifestation.

To summarize, we have presented a model of a non-Maxwellian field that breaks time inversion symmetry, which can account for recent experimental observations, and establish a direct link between non-Maxwellian fields and Berry's curvature. Perturbatively, in second order, and with the crucial assist of spin-orbit coupling, circularly polarized light either lifts Kramers' degeneracy on its own or does so by transferring rotational coherence to the vector field associated with an electric-dipole-allowed excitation of the system. As in the conventional Faraday effect, the static field can rotate the polarization of a linearly polarized probe beam that propagates through the sample. Because the source of the non-Maxwellian field is not magnetic, the equivalent magnetic fields are significantly larger than those due to the IFE.

FIGURE CAPTION

FIGURE 1: Energy diagram showing light-induced lifting of Kramers' degeneracy for a system with a *p*-type (*J* = 3/2) ground state and excited states of *s*- and *p*-type (*J* =1/2) character. Not to scale, the levels simulate the zone-center top valence and lowest-lying conduction band states of a zincblende semiconductor such as GaAs. The perturbation is $e[r_-\mathcal{E}_L(\Omega) + z\mathcal{E}_z(\Omega)] + c.c.$ Dashed lines indicate allowed transitions. To second order, the rectified field mixes, separately, the $|1/2, \pm 1/2\rangle$ and the $|0, \uparrow\downarrow\rangle$ states. The effective mixing Hamiltonian, Eq. (8), is proportional to $a_0^3 e^2 \hbar^{-1}(\mathcal{E}_L \mathcal{E}_z^* - \mathcal{E}_z \mathcal{E}_L^*)/(\Delta^2 - \Omega^2)$ where $\hbar\Delta$ is the energy difference between the *s* and the $|1/2, \pm 1/2\rangle$ states and $\Omega$ is the pump frequency. The symmetry breaking field is equivalent to a magnetic field along the *x*-direction.

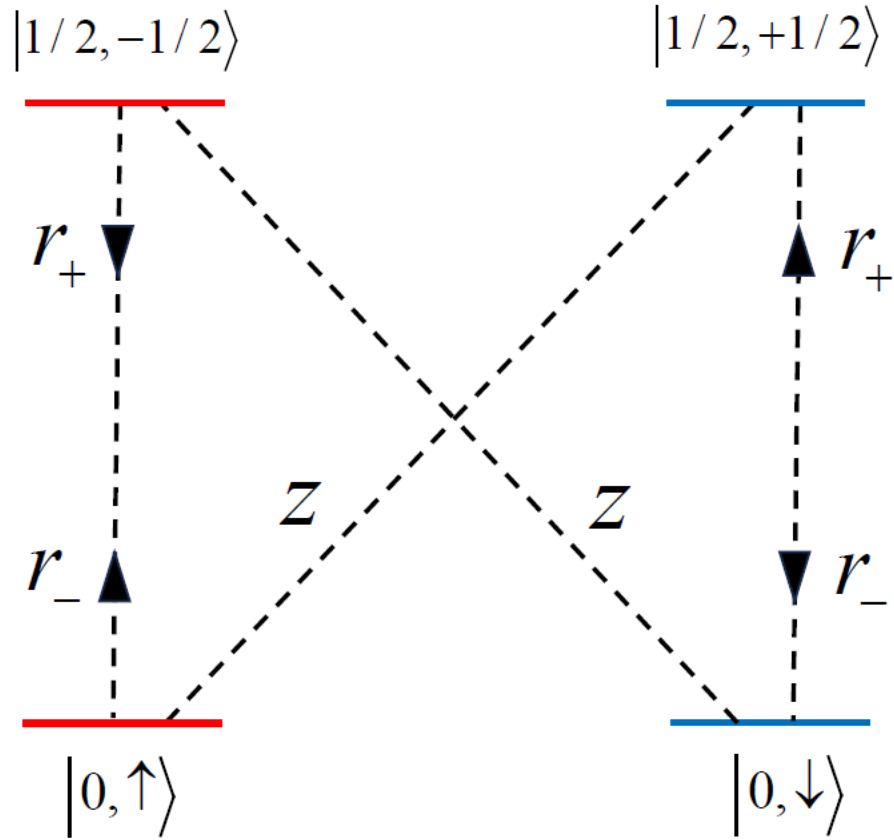
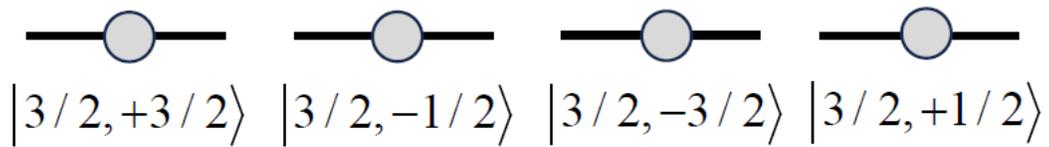